\documentclass[mathleft
]{an}
\usepackage{graphicx}
\usepackage{times}
\newcommand{\lsim}{\ \raise -2.truept\hbox{\rlap{\hbox{$\sim$}}\raise5.truept
       \hbox{$<$}\ }}
\newcommand{\gsim}{\ \raise -2.truept\hbox{\rlap{\hbox{$\sim$}}\raise5.truept
       \hbox{$>$}\ }} 

\begin{document}

% The following seven commands are intended for editorial usage and should be ignored by
% the author(s).
\Pagespan{1}{}% Document's page range. 
% If second parameter is left empty, the last page is computed automatically.
\Yearpublication{2006}%
\Yearsubmission{2005}%
\Month{11}%   
\Volume{999}%  
\Issue{88}% 
% \DOI{This.is/not.aDOI}% 

\title{Gamma-ray burst host galaxies at low and high redshift}

\author{Sandra Savaglio\inst{1}\fnmsep\thanks{Corresponding author:
  \email{savaglio@mpe.mpg.de}\newline}
%Example 
%for footnote, note the usage of the \texttt{fnmsep}
%command as separator between institute number and footnote mark} 
}
\titlerunning{GRB Host Galaxies}
\authorrunning{S.\ Savaglio}
\institute{
Max Planck Institute for extraterrestrial Physics, PO Box 1312, Giessenbachstr., 85741 Garching, Germany}

\received{15 Jan 2012}
%\accepted{?? ?? 2012}
%\publonline{later}


\abstract{The galaxies hosting the most energetic explosions in the universe, the gamma-ray bursts (GRBs), are generally found to be low-mass, metal poor, blue and star forming galaxies. However, the majority of the targets investigated so far (less than 100) are at relatively low redshift, $z<2$. We know that at low redshift, the cosmic star formation is predominantly in small galaxies. Therefore, at low redshift, long-duration GRBs, which are associated with massive stars, are expected to be in small galaxies. Preliminary investigations of the stellar mass function of $z<1.5$ GRB hosts does not indicate that these galaxies are different from the general population of nearby star-forming galaxies. At high-$z$, it is still unclear whether GRB hosts are different. Recent results indicate that a fraction of them might be associated with dusty regions in massive galaxies. Remarkable is the a super-solar metallicity measured in the interstellar medium of a $z=3.57$ GRB host.}

\maketitle

\section{Introduction}

Gamma-ray bursts (GRBs) are the most energetic explosions in the universe. Discovered in 1967 by Vela satellites\footnote{This was a US Defense Department mission aiming at spying nuclear tests during the cold war.}, they are associated with the death of the most massive stars (supernovae) or mergers of compact stellar objects (neutron-stars or black-holes). The reason for the relatively recent discovery is the short-lasting (if at all observed) optical emission of the afterglow, generally less than a few weeks. More than 40 years of controversies is well represented by an emblematic event: for GRB 060218, detected on 2006 February 18, more than 20 papers were written in two years, four of which appeared in the famous science journal {\it Nature}. 

The typical energy radiated (in photons) by a GRB is 10$^{51}$ ergs, mainly in $\gamma$-rays, in a time interval that often does not exceed one minute. This is equivalent to the energy emitted by Sun over its entire life (more than 10 billion years). Other forms of energy, some dominating, are not easily detectable by current technologies. Gravitational waves can emit 10$^{49}$ ergs, the kinetic energy in the explosion is 10$^{53}$ ergs, and the energy carried by the neutrinos is at least ten times more. Being so energetic, life would not be possible on Earth if they were living too long or too common. It is estimated that only one supernova (SN) over 1000 turns into a GRB, and the rate in a normal galaxy is 1/10$^5$ yr$^{-1}$. Nevertheless, if one considers all galaxies, a few events every day are in principle detectable by  $\gamma$-ray instruments in space.

In the latest years, a sequence of extraordinary discoveries helped not only to understand their nature, but also to explore the universe under extreme conditions. In March 2008, NASA satellite {\it Swift} detected  GRB\,080319B (Bloom et al.\ 2009), the optically brightest event ever recorded by humanity. At a relatively large distance ($z = 0.937$, exploded when the universe was 45\% of its present age), the optical afterglow had a peak magnitude $m = 5.6$. The following year, a few very distant bursts were observed: first GRB\,080913  at $z = 6.695$ (Greiner et al.\ 2009), then GRB\,090423 at $z = 8.3$ (Salvaterra et al.\ 2009; Tanvir et al.\ 2009), finally GRB\,090429B at $z \sim 9.4$ (Cucchiara et al.\ 2011). All these events occurred when the universe was younger than 900 Myr. Few months later, the high energy photons of GRB 090510 ($z=0.903$), detected by {\it Fermi Gamma-Ray Space Telescope},  were used to test the Lorentz invariance. Formulated in Einstein's special relativity theory, Lorentz invariance predicts that photons with different energies have the same speed. A 31 GeV photon, emitted 0.829 seconds after the burst, did not show any even tiny speed variation (Abdo et al.\ 2009), disfavoring a number of quantum-gravity theories\footnote{This is true only under the assumption that the GeV photon was emitted at the same time as lower-energy emission.} which predict Lorentz invariance violation below the Planck scale (energy $> 1.22\times10^{19}$ GeV).

GRBs are routinely used to probe a variety of physical phenomena going from general relativity, to black holes, to cosmology. Among these, the investigate of galaxy formation and evolution through GRBs is one of the most controversial. Galaxies hosting GRBs offer the opportunity to explore galaxies that are not easy to detect using traditional means. A GRB is a flash of $\gamma$-rays, whose detection is not affected by the presence of dust along the sightline or the brightness of the parent galaxy. Here we briefly report about accepted facts, latest results, and issues that are still disputed about this population of nearby and distant galaxies.

\begin{figure*}
\includegraphics[width=110mm,height=60mm]{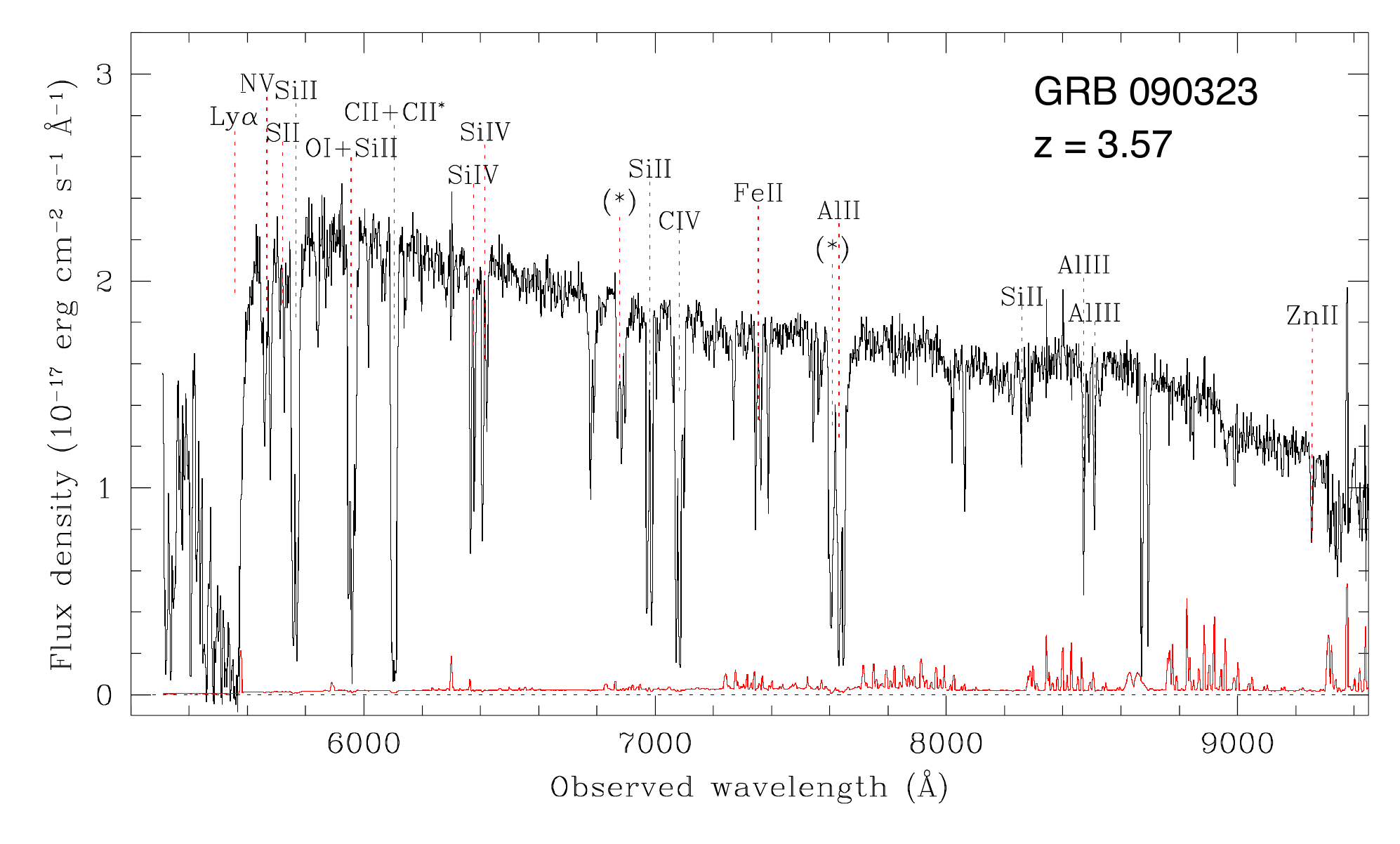}
\includegraphics[width=61mm,height=57mm]{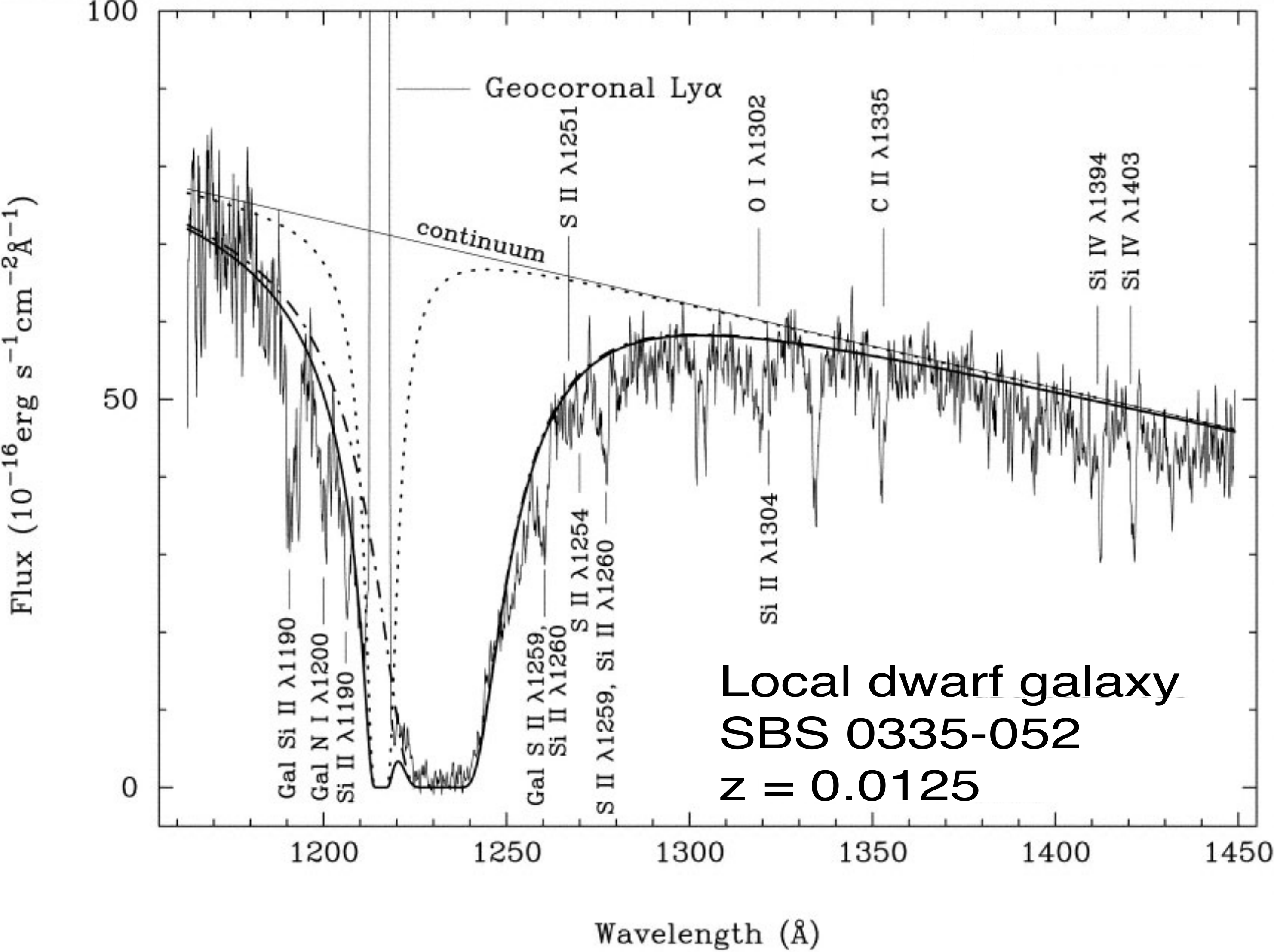}
\includegraphics[width=114mm,height=50mm]{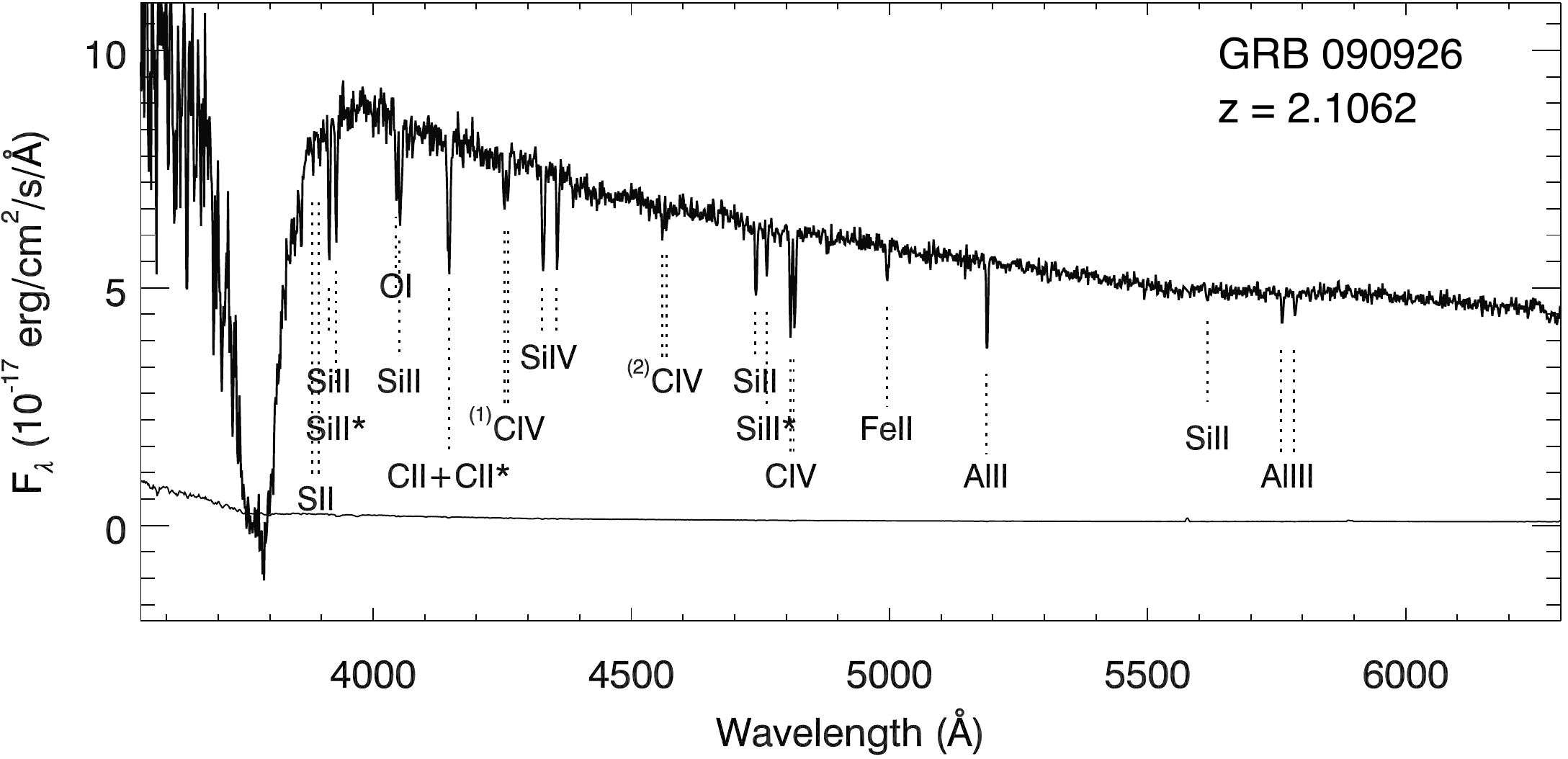}
\caption{Comparison between two high-$z$ GRB afterglow spectra (probing the cold ISM in the host galaxy) and the spectrum of a local dwarf star-forming galaxy. {\it Upper-left panel}: spectrum of the afterglow of GRB\,090323 at $z=3.57$ (Savaglio et al.\ 2011), with strong metal absorbers revealing  the presence two systems separated by 660 km s$^{-1}$ with high (super solar) metallicity. {\it Lower-left panel}: spectrum of the afterglow of GRB\,090926 at $z=2.1062$ (Rau et al.\ 2010), characterized by a strong H\,{\sc i} absorption and weak metal lines (low metallicity). {\it Upper-right panel}: spectrum of the local ($z=0.0125$) dwarf star-forming galaxy SBS 0335-052 (Thuan \& Izotov 1997). This galaxy has high H\,{\sc i} content and weak metal lines, very similar to the spectrum of GRB\,090926, and different from that of GRB\,090323. The presence of the fine-structure lines of the singly-ionized silicon Si\,{\sc ii}* is common to the ISM of all three galaxies.}
\label{fig1}
\end{figure*}

\begin{table*}
\centering%%%
\caption{Galaxy parameters in a local dwarf galaxy and GRB host galaxies}
\label{tab1}
\begin{tabular}{l|cccc}\hline
Parameter & SBS 0335-052 & GRB 980425 & GRB 090926B & GRB 930323 \\ 
\hline
\hline
Redshift			& 0.0125 & 0.0085 & 2.1062 & 3.57 \\
$M_B$ 			& $-16.9$ & $-18.6$ & -- & $-24.9$ \\
Size (kpc$^2$) 		& $6\times5$ & $30\times20$ & -- & $<6\times6$ \\
$\log Z/Z_\odot$ 	& $-1.4$ & $-0.5$ & $-1.9$ & $+0.25$ \\
$N_{\rm H\,I}$ (cm$^{-2}$) & $7\times10^{21}$ & -- & $5.4\times10^{21}$ & $5.6\times10^{20}$ \\
$M_{\rm H\,I}$ (M$_\odot$) & $\sim8\times10^8$ & -- & --  & -- \\
$M_\ast$ (M$_\odot$) & $\sim4\times10^7$ & $\sim2\times10^9$ & -- & $\sim6\times10^{10}$  \\
SFR (M$_\odot$ yr$^{-1}$) & 0.5 & 0.2 &  -- & $>6$ \\
SSFR (Gyr$^{-1}$) & 12.5 & 0.1 & -- & $>0.1$ \\
Age (Myr) & $<400$ & $\sim 900$ &  --  & $<500$ \\
\hline
\end{tabular}
\end{table*}

\section{GRB host galaxies}

For only about half of all GRBs with known redshift (more than 240), the hosting galaxy is studied in some detail. Most of our knowledge is based on the observations of those at $z<2.3$ (87 galaxies, 71\% of the total). In the past, it was shown that most of them are small, star-forming, blue and metal-poor galaxies (Vreeswijk et al.\ 2001; Christensen et al.\ 2004; Fruchter et al.\ 2006; Wiersema et al.\ 2007). However, numerous new and deeper observations suggest that this might be a partial view, probably affected by a combination of the difficulty of detecting distant targets and the redshift evolution of galaxy fundamental parameters (Kr{\"u}hler et al.\ 2011; Savaglio et al. 2011). What is concluded from low-$z$ explorations might not be valid at high redshift.

Fig.\,\ref{fig1} shows that the spectra of high-$z$ GRB afterglows, probing the cold ISM in the host galaxy, are not necessarily similar to the spectrum of a typical local dwarf galaxy. The host of GRB\,090926A at $z=2.1062$, being metal poor and gas rich (Rau et al.\ 2010), indeed resambles the $z=0.0125$ dwarf star-forming galaxy SBS 0335-052 (Thuan \& Izotov 1997). However, the host of the distant GRB\,090323 ($z=3.57$) has a peculiar strong double absorption system with super-solar metallicity (Savaglio et al.\ 2011). The properties of the ISM in these galaxies are summarized in Table\,\ref{tab1}. GRB afterglow spectra are often characterized by the presence of excited lines of silicon S\,{\sc ii}*, indicating the vicinity of a bright UV source. It is true that this species was never detected in the ISM probed by absorption lines in QSO spectra (damped Lyman-$\alpha$ systems, DLAs). However, it is detected in SBS 1335-052 (Thuan \& Izotov 1997) and in Lyman-break galaxies (Pettini et al.\ 2002), suggesting that Si\,{\sc ii}* is not necessarily identifying a GRB explosion in the vicinity, but common in regions of intense star formation. This is further proven by the fact that S\,{\sc ii}* is present in both the absorbers detected in the GRB\,090323 afterglow spectrum. The separation of the two absorbers is 660 km s$^{-1}$ (Savaglio et al.\ 2011), the GRB can be close to one of the two, but not both. 

\begin{figure}
\includegraphics[width=82mm]{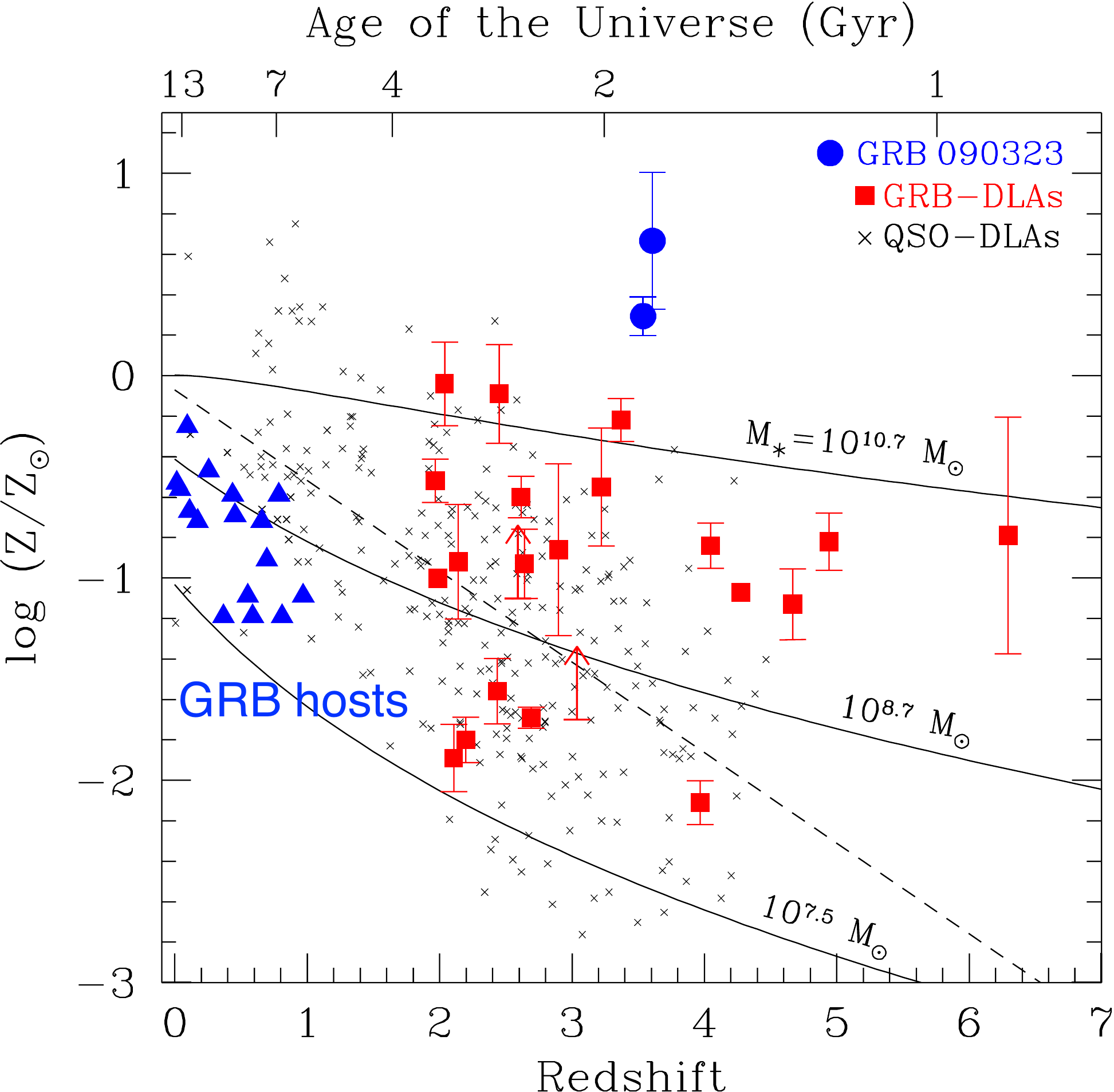}
\caption{Metallicity as a function of redshift in galaxies. Metallicities in the two absorbers detected in GRB\,090323 (blue dots) are separated by 660 km s$^{-1}$. Red squares are metallicities of other GRB-DLAs. Blue triangles are GRB host metallicities measured from emission lines at low redshift (Savaglio et al.\ 2009). Metallicities in other high-$z$ absorbers detected in QSO spectra (QSO-DLAs) are black crosses (see Savaglio et al.\ 2011 for more details). The dashed line is the linear correlation for QSO-DLA points. Solid curves are metallicities expected for star-forming galaxies with different stellar masses (from the redshift evolution of mass-metallicity relation; Savaglio et al.\ 2005).}
\label{fig2}
\end{figure}

In Table\,\ref{tab1}, we include the properties of the host of the well-known GRB\,980425 at  $z=0.0085$ (Micha{\l}owski et al.\ 2009; Savaglio, Glazebrook \& Le Borgne 2009). The global properties of the three GRB hosts have little in common with the dwarf galaxy SBS 1335-052 (Pustilnik et al.\ 2004). The question on what is the galaxy parameter triggering a GRB event is still not answered. It seems that low metallicity, as typically assumed, is not a valid requirement, as shown by the high metallicity of GRB\,090323. Metallicities, derived from absorption lines in GRB afterglow spectra (cold ISM) and emission lines in integrated spectra of GRB hosts (warm ISM), as a function of redshift, shows a large dispersion (Fig.\,\ref{fig2})  with relatively low values at low redshift, and relatively high values at high redshift.

High SFR is often detected in GRB hosts. The progenitor of a long-duration GRB is a massive and short-lived star ($M > 30$ M$_\odot$; Heger et al.\ 2003). Similar stars can only be located in regions with intense star formation. If we consider that cosmic star formation rate was higher in the past than it is today (it dropped by a factor of 50 from $z \sim 1.8$; Hayes et al.\ 2010), and that it transited from large galaxies in the past to small galaxies today (Juneau et al.\ 2005), it is natural to think that in the local universe most GRBs occurred in small galaxies, while at high redshift more massive galaxies likely hosted a large fraction of events. The next goal would be to study the mass function of GRB hosts. At the moment, this is vaguely possible, if at all, at low redshift only, due to the small number statistics.

\section{Stellar mass function of GRB host galaxies}

The investigation of the mass function (MF) of galaxies is a fundamental mean through which the cosmic change of galaxies can be identified and understood. The MF of the stellar-mass component of galaxies has been widely investigated in the local universe (e.g., Baldry et al.\ 2008), and now also possible at high redshift (Santini et al.\ 2011). For star-forming galaxies (Pozzetti et al.\ 2010; Gilbank et al.\ 2011), the shape of the MF did not change much over the last 6 Gyr ($z\lsim1$). The difference is the normalization: today we have $\sim3$ times more galaxies with stellar masses $M_\ast > 3\times10^{8}$ M$_\odot$ than back then.

The MF was never derived for galaxies hosting GRBs for two major reasons: the small number statistics and the difficulty in defining the sample completeness. Despite the small sample, the identification of GRB hosts, which is, to first order, independent of the galaxy brightness, makes the investigation of the $z>0$ MF in the low stellar-mass regime ($M_\ast <10^{10}$ M$_\odot$) a possible task.

It is observationally confirmed that GRBs are associated with regions of intense star formation. It is also now known that small galaxies dominate when considering the star-formation density in the local universe,  whle the contribution from large galaxies was more important at high redshift than today (Juneau et al.\ 2005; Papovich et al.\ 2006). Many GRB hosts are indeed small galaxies, but this has been demonstrated in the low-$z$ universe only. It is true that small galaxies are the most common galaxies in the universe (Baldry et al.\ 2008). But, if this is true in the local universe, it must be even more so at $z>2$ (Santini et al.\ 2011): galaxies can only get larger with time, never smaller. Therefore, it is very hard to predict what is the typical galaxy associated with high-$z$ GRBs, all possibilities are open.

It is now important to establish whether GRB hosts at low redshift belong to a unique population, or naturally fill the low-mass end of the galaxy MF. Savaglio et al.\ (2009) studied a sample of 45 GRB hosts and no evidence for deviation from normal galaxies was found. On the other hand, Levesque et al.\ (2010) showed that the mass-metallicity relation in GRB hosts is shifted towards lower metallicities with respect to field galaxies. However, it is not clear if this is due to a generally higher SFR in GRB hosts, as also recently indicated by field galaxies (Mannucci et al.\ 2010).
 
In our ongoing work (Savaglio, Glazebrook, Le Borgne et al., in prep.), the GRB host sample is 2.5 times larger than in Savaglio et al.\ (2009). To investigate the GRB host MF, one needs to select the sample according to the volume limited criteria. This is the most critical part, which would be relatively under control, if we could use a large sample. In general, in any magnitude-limited galaxy sample, the mass distribution is dominated by massive galaxies, stellar mass above $M^*$ galaxies.  When one corrects for volume, then low-mass galaxies dominate. For GRB hosts, the sample selection is totally different.  If 100\% of detected GRBs were followed up and galaxy masses are measured, then part of the relevant volume calculation would be the flux limit of the GRB-detecting telescopes (i.e., {\it Swift}) as a function of redshift. The bias in favor of certain kind of galaxies, e.g., star forming, is at this point a detail.

\begin{figure}
\includegraphics[width=83mm]{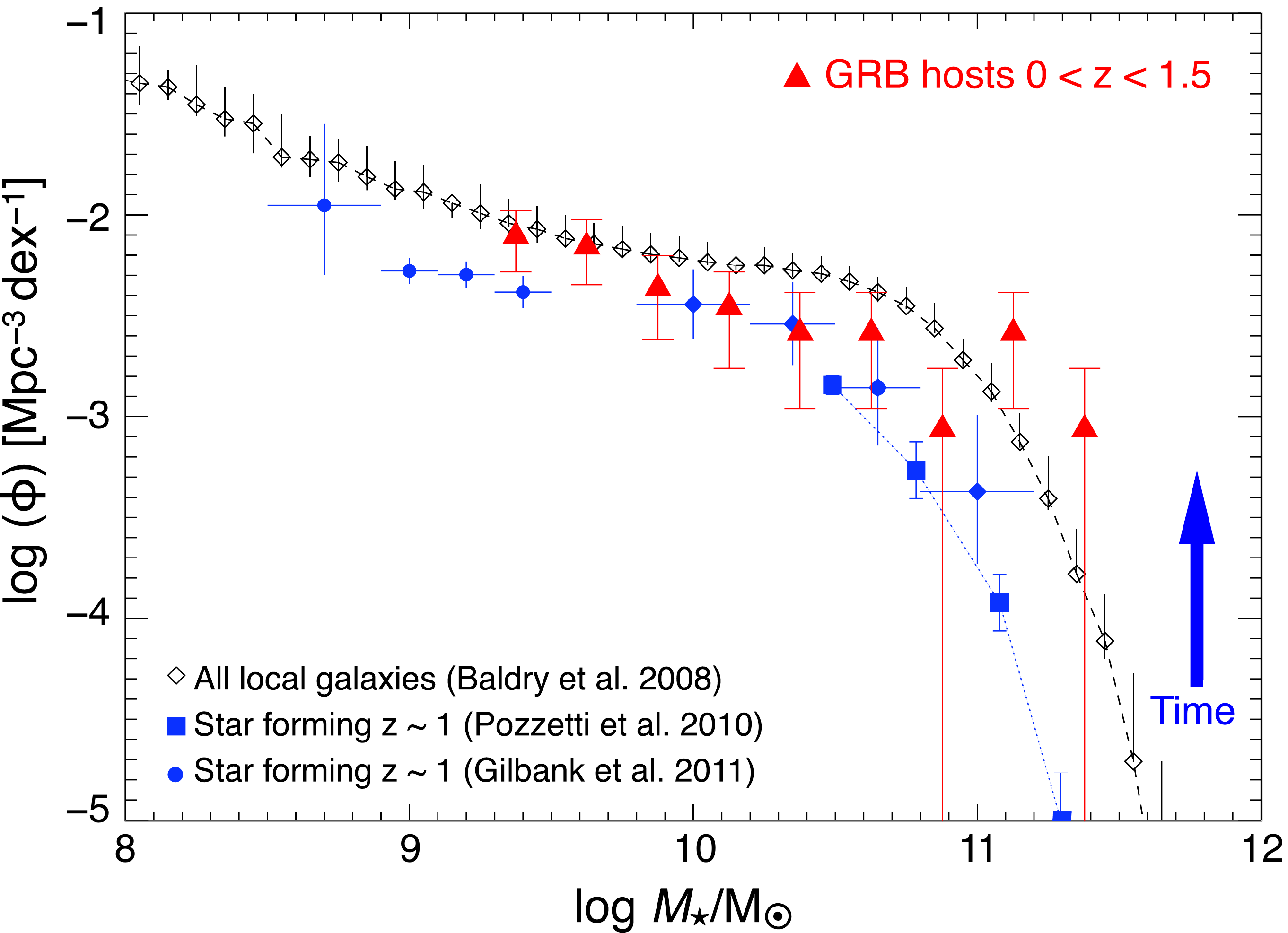}
\caption{Stellar mass function (MF) for field galaxies and GRB hosts at different redshifts. Black diamonds is the MF for all $z\sim0$ galaxies (Baldry et al.\ 2008), blue squares and filled circles is the $z \sim 1$ MF for star-forming galaxies, from Pozzetti et al.\ (2010) and Gilbank et al.\ (2011), respectively. Red triangles is the MF estimated from a sample of 31 GRB hosts in $z < 1.5$ and $\log M_\ast/{\rm M}_\odot > 9.25$, and normalized to be in between field-galaxy MFs in the interval $9.25 < \log M_\ast/{\rm M}_\odot < 10.70$ (Savaglio, Glazebrook, Le Borgne et al., in prep.).}
\label{fig3}
\end{figure}

As a preliminary approach, we skip the correction functions and investigate just the shape of the observed MF, by normalizing it to that of field galaxies.  We apply to our sample a mass cut (mass completeness limit) instead of luminosity cut, as typically done, because the UV-optical brightness of galaxies can vary by a large factor, regardless of mass, due to different SFR. Fig.\,\ref{fig3} shows the MF  of 31 GRB hosts with $z< 1.5 $ and $M_\ast > 10^{9.25}$ M$_\odot$, and the comparison with field galaxies in the local universe (Baldry et al.\ 2008) and star-forming galaxies at $z\sim1$ (Pozzetti et al.\ 2010; Gilbank et al.\ 2011). The comparison indicates that the general shape is close to that of field galaxies. GRB hosts have a reputation of being small star-forming galaxies, but this is not apparent from this relatively flat MF. 

The most important issue to solve now is the primary selection function of {\it Swift} in a luminosity vs.\ redshift plot. We will have to determine whether the actual luminosity distribution of GRBs (in $\gamma$-rays) is a gaussian, a power law, or a Schechter luminosity function. Once this is under control, we can apply the volume-limited calculation.

\section{A multi-wavelength approach}

The  radiation emitted by the stellar component of galaxies dominates the spectral energy distribution (SED) for wavelength $< 4\, \mu$m. Molecular gas and dust, which are fundamental ingredients for star formation in young galaxies, are heat by young stars and re-emit for longer wavelength. This part of the SED is extremely informative of the history and physical state of galaxies, but observationally very challenging. Given that the global star-formation rate dropped by a factor of $\sim 50$ from $z \sim1.8$ (Hayes et al.\ 2010), molecular gas and dust were a much more important component at high redshift than today. The investigation of GRB hosts beyond the near IR can reveal, on that regard, surprising results.

As discussed above, GRB hosts at high redshift might be much more diverse than at $z<2$. High-$z$ observations  are often limited to optically bright GRB afterglows, for which the redshift can actually be measured. Dark GRBs, those for which the optical afterglow emission is very faint relative to the extrapolation from the X-ray (Jakobsson et al.\ 2004; van der Horst et al.\ 2009), can either be objects at $z>7$ or embedded in regions with high dust content. It was recently found that a large fraction of dark GRBs are in massive, star-forming galaxies with red colors, high extinction and large SFRs (Berger et al.\ 2007; Chen et al. 2010; Hunt et al.\ 2011; Greiner et al.\ 2011). Once more, the assumption that all GRB hosts are low-mass, dust free galaxies may be an oversimplification.

The statistics of dark GRBs and their host galaxies is still poor, because the optical faintness makes the localization very difficult. However, it was possible to confirm that in most dark GRBs, the optical faintness is caused by high dust extinction columns and moderate redshift (Perley et al.\ 2009; Zafar et al.\ 2011). Some GRB hosts are blue, but  a significant amount of dust with patchy distribution can explain the red color of the afterglow (Kr\"uhler et al.\ 2011). The situation is complex. While it is true that often massive, metal-rich GRB hosts are found from dark GRBs, not all dark GRBs reside in massive hosts, nor are they all found at high redshift. Direct observations of the entire SED would help to place them in the context of other high-$z$ galaxy populations, although also most of what we know about high-$z$ galaxies is based on rest-frame UV observations. So far, only a handful of GRB hosts have been detected with sub-millimeter (submm) facilities (Tanvir et al.\ 2004; Micha{\l}owski et al.\ 2008;  Micha{\l}owski et al.\ 2009). In this minority, SFRs can be high, $\sim 500$ M$_\odot$ yr$^{-1}$, as high as those of the submm galaxy (SMG) population (Blain et al.\ 1999; Ivison et al.\ 2002). There has been no CO emission found in any of the GRB hosts observed so far (Hatsukade et al.\ 2011), and there is some hint that dust in GRB hosts may be warmer than in typical SMGs (Priddey et al.\ 2006).

\begin{figure}
\includegraphics[width=83mm]{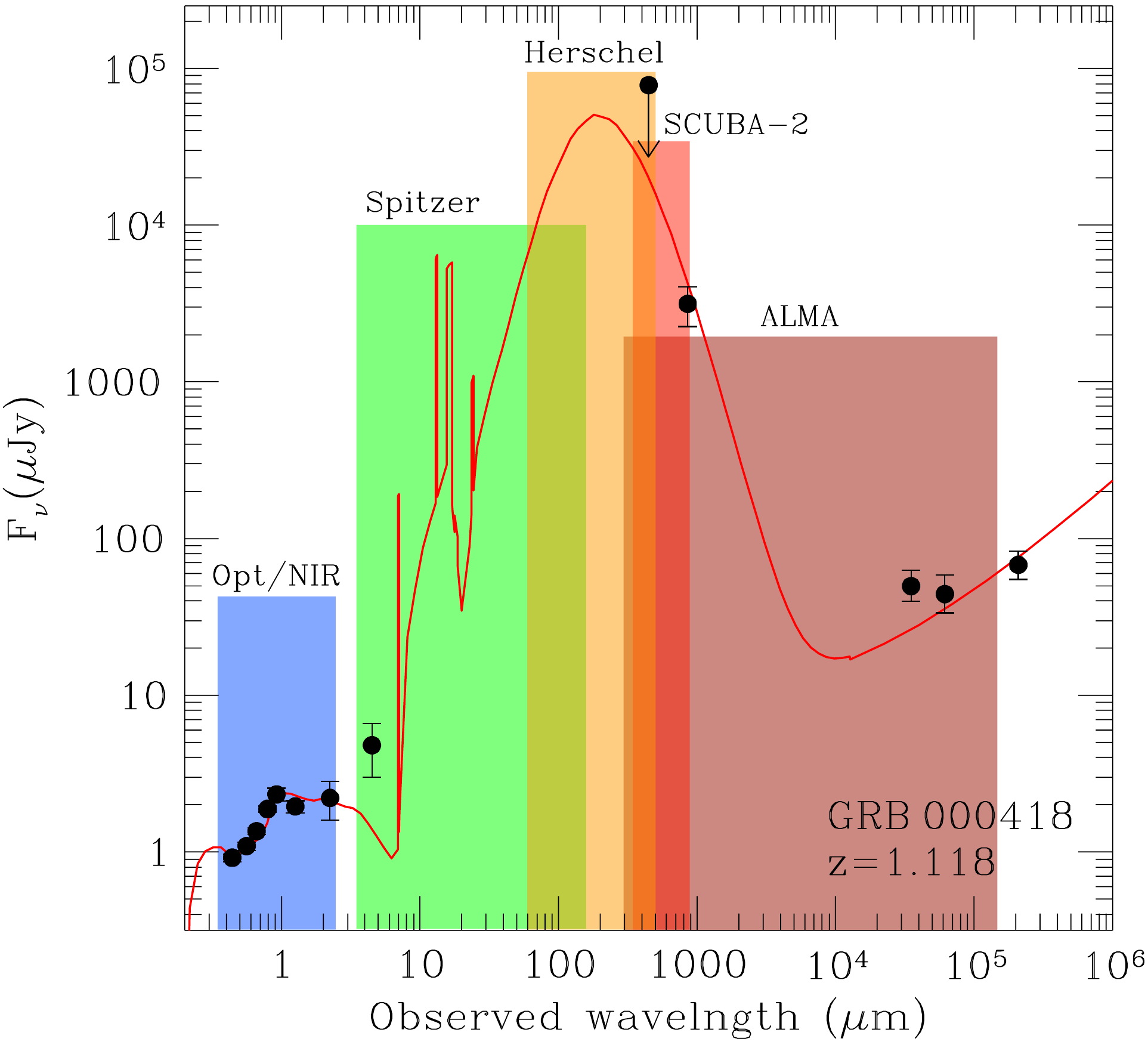}
\caption{Spectral energy distribution of a typical GRB host galaxy, the host of GRB\,000418 at $z=1.118$. Data (filled dots) are taken from Berger et al.\ (2003), Gorosabel et al.\ (2003), Micha{\l}owski et al.\ (2008), and Castro Cer\'on et al. (2010). The red curve is a young starburst galaxy model calculated using GRASIL (Micha{\l}owski et al.\ 2008). Shaded areas show the coverage of present FIR, submm and radio telescopes.}
\label{fig4}
\end{figure}

The exploration of GRB hosts at long wavelengths is still relatively recent. Targets are distant and faint. One well known case is the host of GRB\,000418 at $z=1.118$, whose SED is shown in Fig.\,\ref{fig4} (Berger et al.\ 2003; Micha{\l}owski et al.\ 2008; Castro Cer{\'o}n et al.\ 2010). The host was exceptionally bright in the FIR-radio and was detected with {\it Spitzer}, {\it SCUBA} and {\it VLA}. Fig.\,\ref{fig4} shows also the coverage of some of the present instrumentations ({\it Spitzer}, {\it Herschel}, {\it SCUBA-2}, and {\it ALMA}). They are our main tools to unveil the gas and dust properties for the largest possible sample of GRB hosts.

\section{Summary}

Galaxies hosting GRBs are important probes of galaxy formation and evolution, because they are selected according to criteria that are very different from those used for traditional galaxy surveys. Thanks to the GRB event, it is possible to explore regions of the universe and galaxies that are too dusty, too faint, or too far to be seen with optical and NIR instruments. For a long time, GRB hosts have been accused of belonging to a biased sample, thus not properly representing the general galaxy population. Most of them are indeed associated with low-mass, low luminosity, metal poor, dust free galaxies, similar to local blue dwarf galaxies. However, similar galaxies are the most common galaxies that are forming stars in the nearby universe ($z<2$). Star formation is required to trigger a long-duration GRB, whose progenitor is a massive star. All this is not particularly surprising.

On the other hand, what is the typical GRB host at $z>2$ is less clear. The cosmic chemical evolution derived with GRBs is in that regard telling something important. The absorption lines in the afterglow spectra were used to probe the metal enrichment in the cold ISM for $\sim20$ host galaxies at $z>2$. The metallicity shows a large spread of more than two orders of magnitude, with no trend with redshift. Striking is the recent super-solar metallicity measured in the $z=3.57$ GRB\,090323, and the universe at that time was only 1.7 Gyr old.

We argue that low metallicity is perhaps not the main ingredient generating a GRB event, not at high redshift at least. Star formation activity seems to be a more stringent requirement. The cosmic SFR is a strong function of redshift, meaning that at low redshift it is happening in small galaxies, whereas at high redshift it does not seem to be the case anymore. Therefore, GRB hosts at high redshift might show a larger variety of stellar mass than at low redshift. Deriving the stellar mass function for GRB hosts is complicated by the particular selection function and the small number statistics. From a preliminary attempt, we conclude that the shape of the mass function for a small sample of $z<1.5$ GRB hosts with $M_\ast > 10^{9.25}$ M$_\odot$ is consistent with the mass function of field galaxies.

Our on-going effort is to systematically investigate a large sample of GRB hosts at long wavelengths, using new instruments and telescopes (e.g., {\it ALMA}).  The multiwavelength approach will show whether a fraction of GRB hosts at high redshifts is similar to submm galaxies, characterized by high SFR, dust, gas, metallicity, and merging rate. This would demonstrate that the population of GRB hosts at high redshift is very different from the nearby population.

Most GRBs discovered these days rely on the performance of the dedicated mission {\it Swift}. This NASA satellite is one of the most successful telescopes in operation in general, which is remarkable, given its medium-size budget. {\it Swift} will be supported until 2014, and is right now under evaluation for extended support for the following years. The only other mission foreseen for the near future is the satellite {\it SVOM}, a joint project of the Chinese National Space Agency and the French Space Agency, to be launched after 2015.

\acknowledgements
I express my appreciation to St{\'e}phane Basa, Karl Glazebrook, Jochen Greiner, Leslie Hunt, Damien Le Borgne, Michal Micha{\l}owski and Eliana Palazzi for the fruitful collaboration.

%\newpage%%%%%%%%%%%%%%%%%%%%%%%%%%%%%%%%%%%%%%%%%%%%%%%%%%%%%%

\end{document}